\documentclass[sigconf,screen=true,bookmarks=false, 9pt]{acmart}
\usepackage{algorithm}
\usepackage{graphicx}
\usepackage{threeparttable}
\usepackage{color}
\usepackage{colortbl}
\usepackage{multirow}
\usepackage{amsmath}
\usepackage{comment}
\usepackage{bm}
\usepackage{etoolbox}                                      
\usepackage{bbm}
\usepackage{url}
\usepackage{nth}
\usepackage{balance}
\usepackage{courier}                                       
\usepackage{booktabs}                                      
\usepackage{listings}                                      
\usepackage[]{algpseudocode}
\usepackage{xspace}

\algdef{SE}[DOWHILE]{Do}{doWhile}{\algorithmicdo}[1]{\algorithmicwhile\ #1}%

\newcommand{\JG}[1]{\textcolor{blue}{[JG: #1]}}
\DeclareMathOperator*{\argmin}{argmin}

\newcommand{\name}{\texttt{SuperMixer}\xspace}

\usepackage[subrefformat=parens,labelformat=parens]{subfig}
\graphicspath{{./figs/}}

\usepackage{tikz}
\usepackage{bm}
\usetikzlibrary{arrows.meta}

\definecolor{C0}{HTML}{66c2a5}
\definecolor{C1}{HTML}{fc8d62}
\definecolor{C2}{HTML}{8da0cb}
\definecolor{C3}{HTML}{e78ac3}
\definecolor{C4}{HTML}{a6d854}
\definecolor{C5}{HTML}{ffd92f}
\definecolor{C6}{HTML}{e5c494}

\copyrightyear{2022}
\acmYear{2022}
\setcopyright{acmcopyright}\acmConference[ICCAD '22]{IEEE/ACM International Conference on Computer-Aided Design}{October 30-November 3, 2022}{San Diego, CA, USA}
\acmBooktitle{IEEE/ACM International Conference on Computer-Aided Design (ICCAD '22), October 30-November 3, 2022, San Diego, CA, USA}
\acmPrice{15.00}
\acmDOI{10.1145/3508352.3549449}
\acmISBN{978-1-4503-9217-4/22/10}

\begin{document}

\settopmatter{printacmref=False} 
\pagestyle{plain} 

\title{
Fuse and Mix: MACAM-Enabled Analog Activation for Energy-Efficient Neural Acceleration
}

\author{Hanqing Zhu\textsuperscript{$\star$}, Keren Zhu, Jiaqi Gu, Harrison Jin, \\ Ray T. Chen, Jean Anne C. Incorvia, David Z. Pan\textsuperscript{$\diamond$}
\\
ECE Department, The University of Texas at Austin, Austin, TX, USA \\
\textsuperscript{$\star$}hqzhu@utexas.edu; \textsuperscript{$\diamond$}dpan@ece.utexas.edu
}

\begin{abstract}
\label{abstract}
Analog computing has been recognized as a promising low-power alternative to digital counterparts for neural network acceleration. 
However, conventional analog computing is mainly in a mixed-signal manner.
Tedious analog/digital~(A/D) conversion cost significantly limits the overall system's energy efficiency.
In this work, we devise an efficient analog activation unit with magnetic
tunnel junction (MTJ)-based analog content-addressable memory (MACAM), simultaneously realizing nonlinear activation and A/D conversion in a \textit{fused} fashion.
To compensate for the nascent and therefore currently limited representation capability of MACAM,
we propose to \textit{mix} our analog activation unit with digital activation dataflow.
A fully differential framework, \name, is developed to search for an optimized activation workload assignment,
adaptive to various activation energy constraints.
The effectiveness of our proposed methods is evaluated on a silicon photonic accelerator.
Compared to standard activation implementation,
our mixed activation system with the searched assignment can achieve competitive accuracy with $>$60\% energy saving on A/D conversion and activation.
\end{abstract}

\maketitle
\section{Introduction}
\label{sec:Introduction}
Deep neural networks~(DNNs) have received an explosion of interest due to state-of-the-art inference accuracy in a myriad of artificial
intelligence tasks.
In parallel, the rapidly escalating model size and data volume raise a surging need for more efficient computing solutions.
However, as Moore's law winds down, it becomes increasingly challenging for conventional digital counterparts to meet the computational demands of DNN workloads.
A slew of new processor architectures employing analog techniques is keenly sought to reduce power dissipation and improve computational speed.
Crossbar-based processing-in-memory (PIM) architectures~\cite{HWA_NatureReviewM2020_Wang, HWA_DAC2020_roy, HWA_Nature2022_Jung, HWA_ISCA2016_Shafiee, HWA_ISCA2016_Chi} and integrated optical neural networks (ONNs)~\cite{NP_NATURE2017_Shen, NP_NaturePhotonics2021_Shastri, NP_arXiv2021_Feng, NP_TCAD2022_Zhu, NP_TCAD2022_Gu, NP_DAC2022_Gu} are two prominent examples in this direction.

\begin{figure}
    \centering
    \vspace{-5pt}
    \subfloat[]{
        \includegraphics[width=0.22\textwidth]{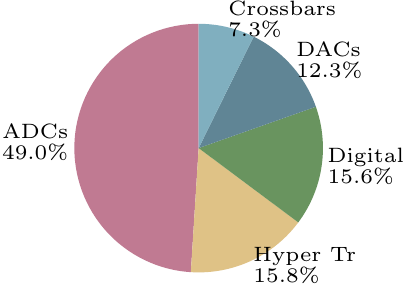}
        \label{fig:IsaacPower}
    }
    \subfloat[]{
        \includegraphics[width=0.22\textwidth]{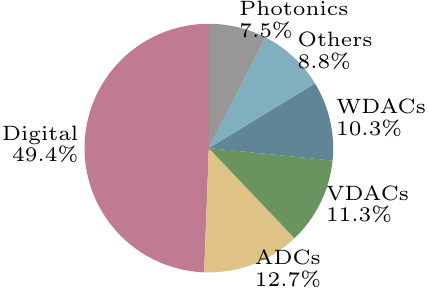}
        \label{fig:LightMatterPower}
    }
    \vspace{-5pt}
    \caption{~\small
    Power breakdown of 
    (a)  a crossbar-based CNN accelerator~\texttt{ISAAC}~\cite{HWA_ISCA2016_Shafiee}
    and (b) a silicon photonic accelerator~\texttt{Mars}~\cite{NP_HotChips2020_Ramey}.
    WDACs means DACs for weights, and VDACs means DACs for vector inputs.
    In ~\texttt{Mars}, it chooses a more costly DAC configuration than the ADC configuration, and most of the power consumption in \textit{digital} part comes from data movement.
    }
    \label{fig:Motivition}
    \vspace{-20pt}
\end{figure}

However, analog computing is mainly in a mixed-signal manner:
digital inputs are transformed into analog signals for computation, and then computing results are converted back to the digital domain for the downstream operations, e.g., activation.
Tedious digital/analog (D/A) and analog/digital (A/D) conversion overhead exists and hinders the overall system energy efficiency~\cite{NP_TCAS22022_Gu, HWA_DAC2021_Song, HWA_DATE2020_Zheng}.
Figure~\ref{fig:Motivition} shows two case studies of energy breakdown in two representative analog convolutional NN (CNN) accelerators,
where the A/D conversion achieved by costly analog-to-digital converters (ADCs) counts for a significant portion of overall power consumption.
The A/D conversion overhead concern is further escalated in ONNs due to its high-speed ADC requirements.

Based on the above analysis, we observe a solid demand for seeking another efficient and low-latency alternative to costly ADCs to bring analog signals back to the digital domain.
A recently proposed memristor-based analog content-addressable memory~(ACAM)~\cite{HWA_NatureCom2020_Li} shows up to be a promising candidate with picosecond-level latency and femtojoule-level energy consumption.
BRAHMS~\cite{HWA_DAC2021_Song} has explored a RRAM-based accelerator to use ACAM to implement nonlinear activation, pooling, and A/D conversion successively, thus eliminating the usage of ADCs. 
However, the separate implementation of the three operations requires routing the analog signals back to ACAM three times, raising a signal noise concern.
Moreover, a severe accuracy drop has been seen in~\cite{HWA_DAC2021_Song} when the precision of A/D conversion implemented by ACAM is insufficient.
These obstacles need to be overcome before viable ACAM can be utilized in analog computing to reduce the A/D overhead.

In this work, we devise an efficient analog activation unit based on MTJ-based ACAM~(MACAM), simultaneously implementing nonlinear activation and A/D conversion in a \textit{fused} manner.
To compensate for the limited representation capability of MACAM, we propose a \textit{mixed} activation system that integrates the proposed low-energy analog activation and the conventional high-precision digital activation datapaths.
With a given activation energy constraint, a \name training flow is proposed to automatically learn how to assign activation workloads on the mixed activation system, aiming at balancing the expressiveness and energy cost.

Our main contributions are as follows,
\begin{itemize}
    \item We propose a novel analog and \textit{mixed} activation system for energy-efficient neural network acceleration.
    \item We devise a \textbf{\textit{fused} analog activation unit} based on MACAM that can \emph{simultaneously} achieve nonlinear activation and A/D conversion with significant energy reduction.
    \item We propose a \textbf{\textit{mixed} activation system} that integrates both analog and digital activation dataflows to balance expressiveness and energy efficiency.
    \item We develop a \textbf{\name framework} to automatically learn how to assign activation workloads on our mixed activation system adaptive to various energy constraints. 
    \item A \textbf{learnable nonlinearity threshold} is proposed
    with an \emph{enhanced training recipe} to boost accuracy under limited MACAM resolution, allowing practical application of very nascent technologies to use their benefits and mitigate their present challenges.
    \item We evaluate our methods on a photonic accelerator. 
    Regarding the energy cost of A/D conversion and activation,
    experiment results show that using fully analog activation units gives $\sim$65\% energy saving, and the searched assignment on the mixed activation system can achieve $>$60$\%$ energy reduction with comparable accuracy.
\end{itemize}

\section{Preliminaries}
\label{sec:Background}
\subsection{Analog Content-addressable Memory}

\begin{figure}
    \centering
    \vspace{-5pt}
    \subfloat[]{
        \includegraphics[width=0.24\textwidth]{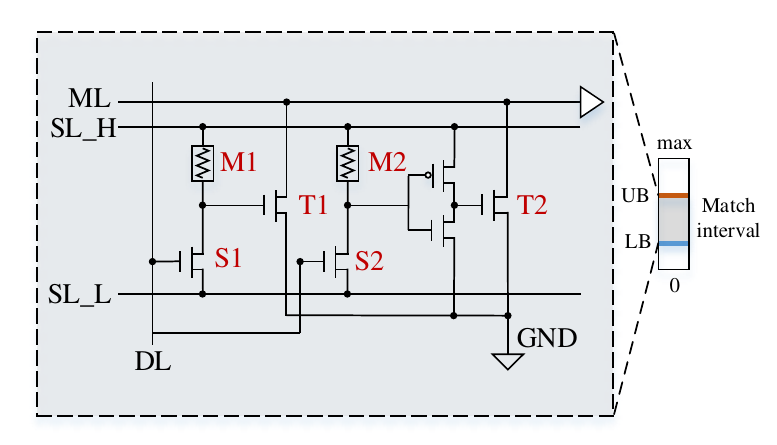}
        \label{fig:schematic_acam}
    }
    \subfloat[]{
        \includegraphics[width=0.20\textwidth]{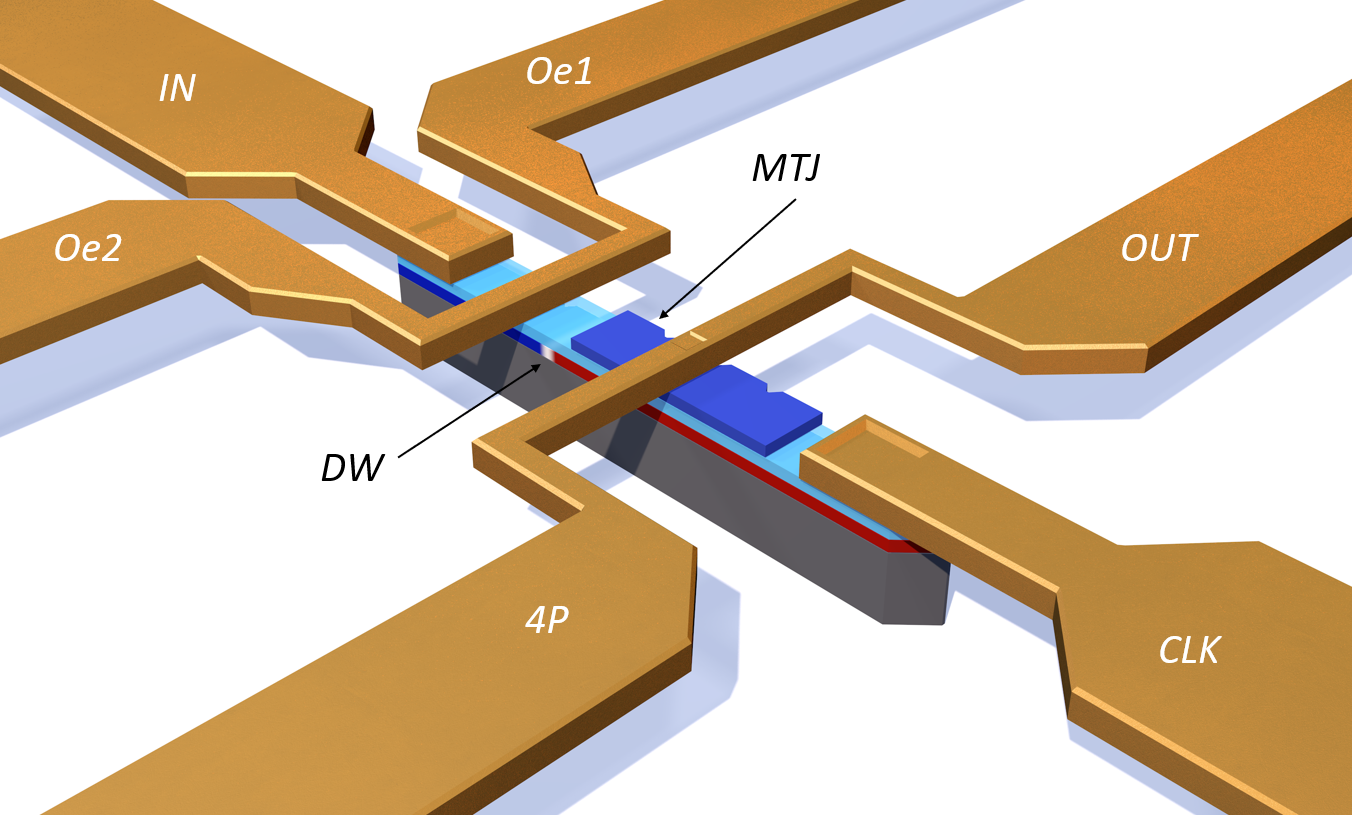}
        \label{fig:schematic_mtj}
    }
    \vspace{-5pt}
    \caption{~\small
    (a) Schematic of the ACAM cell~\cite{HWA_NatureCom2020_Li}.
    (b) The adopted three-terminal domain wall-MTJ~\cite{HWA_Arixv2022_Lenoard}.
    }
    \label{fig:schematic}
    \vspace{-15pt}
\end{figure}

Memristor-based analog content-addressable memory is recently proposed in~\cite{HWA_NatureCom2020_Li}.
Fig.~\ref{fig:schematic_acam} shows the single ACAM cell design that supports search in a continuous analog interval.
The lower bound~(LB) and upper bound~(UB) of the match interval are represented by tuning the resistance of two memristor devices, M1 and M2.
The ACAM cell can take an analog voltage as the input, which is applied to the data line~(DL).
DL is connected to the gate of two switching transistors, S1 and S2.
When the input is smaller (larger) than the LB~(UB), both S1 and S2 are OFF~(ON), making T1~(T2) ON.
Then the match line (ML) will be pulled down, leading to a \emph{mismatch}.
If the input is within the interval, the ML will not be pulled down, resulting in a \emph{match}.
The prior study~\cite{HWA_NatureCom2020_Li, HWA_DAC2021_Song} has proved the ACAM can achieve the search functionality with picosecond-level latency and femtojoule-level energy consumption.
We can cascade multiple ACAM cells to form an ACAM array, where
each cell represents one specific interval such that all intervals can consist of a large search interval.
The maximum search range is decided by the minimum and maximum resistance of the memristor device.
It is obvious that the number of implementable intervals is decided by the number of available resistances of the memristor device.

Regarding the choice of the memristor device, since the ACAM acts as a role of memory with potential frequent access, it inherently requires the chosen memristor to be long-endurance. The
magnetic tunnel junction~(MTJ) turns out to be a more suitable choice with excellent endurance ($10^{15}$ cycles) compared with PCM ($10^7$ cycles) and RRAM ($10^5$ cycles)~\cite{HWA_DAC2020_roy}.
Among MTJs,
Figure~\ref{fig:schematic_mtj} shows
a three-terminal domain wall-MTJ~(DW-MTJ) implementation, which isolates the read (OUT) and write (IN) paths for even better endurance while also providing analog resistance levels with excellent stability of the resistance levels over numerous cycles~\cite{HWA_Arixv2022_Lenoard}.
Hence, in this paper, we choose the DW-MTJ as the memristor device of ACAM.
But, this is a very new MTJ type, and currently the number of resistance levels implemented on DW-MTJ is rather limited, where 5 stable resistance levels are demonstrated in~\cite{HWA_Arixv2022_Lenoard} as the state-of-the-art~(SOTA) implementation.

\subsection{Related Works and Motivation}
Several prior works have explored reducing the A/D conversion overhead in PIM.
In~\cite{HWA_DAC2015_Liu2}, building fully analog circuits is proposed such that the signals are transmitted between layers without A/D conversion.
This fully analog manner has been proved to be accuracy unfriendly.
\texttt{PRIME}~\cite{HWA_ISCA2016_Chi} chooses to use sense amplifiers (SAs) to do A/D conversion instead of ADCs. 
But a SA can only convert one bit at a time, leading to a long latency to get the whole output.
\texttt{CASCADE}~\cite{HWA_MICRO2019_chou} proposes to accumulate the partial sums in the analog domain by connecting the outputs of multiple crossbars via an additional buffer RRAM array. However, the last-mile A/D conversion is kept to convert the sum back to the digital domain for downstream tasks, e.g., activation.

Typically, in analog computing, only matrix multiplications~(MMs) are conducted in the analog domain, while nonlinear activation function and pooling are implemented in the digital domain.
Thus, A/D conversion is necessary.
\texttt{BRAHMS}~\cite{HWA_DAC2021_Song} proposes to put all the MMs, activation, and pooling in the analog domain, where ACAMs implement the latter two in two stages.
In this way, only those analog values which are kept after activation and pooling need to be converted to the digital domain.
The A/D conversion is still done by ACAMs.
In this case, costly A/D conversion can be reduced by using the efficient ACAM.
However, it is prone to signal noise since the analog signals need to be routed through ACAM three times to do nonlinear projection, pooling, and AD conversion.
Moreover, severe accuracy drops are observed when the precision of A/D conversion implemented by ACAM is insufficient.

Therefore, in this paper, we devise a MACAM-based analog activation unit.
It is capable of implementing nonlinear activation and A/D conversion at one time in a fused manner.
It can replace the traditional digital activation path, i.e., costly ADCs and digital activation units.
As a result, the overhead of A/D conversions can be reduced.
Since choosing MTJ as the memristor device will raise the precision issue as well,  a slew of efforts are dedicated to improving the expressiveness of this device type.

\section{MACAM-enabled Fused Analog Activation Unit}
\label{sec:Method}
Nonlinear activation functions in analog computing are normally implemented by digital logics or lookup tables in the digital domain.
Since the computation is conducted in the analog domain, 
analog-to-digital conversion overhead exists to bring analog signals to the digital domain for activation.
We denote this conventional activation implementation as electrical \textit{digital activation}, which includes the needed ADCs and the following digital logics or lookup tables.

Recently,
\texttt{BRAHMS}~\cite{HWA_DAC2021_Song} has explored the idea of using ReRAM-based ACAM to implement nonlinear activation functions and A/D conversions, benefiting from the high-speed possessing and low power consumption of ACAM.
However, the nonlinear projection and A/D conversion are separate.
The analog signals being nonlinearly projected need to be routed back to the ACAM to do the A/D conversion, which may suffer from signal noise.
Moreover, severe accuracy drops are observed when the precision of ACAM-based A/D conversion is not sufficient.

In this paper, we propose to implement the nonlinear activation function using MACAM, where \textbf{nonlinear activation} and \textbf{A/D conversion} are \emph{simultaneously} achieved by the same MACAM array in a \emph{fused} fashion.
The fundamental idea here is to utilize the analog search functionality of MACAM to digitize the analog input signals while introducing \emph{in-situ} nonlinearity.
We call this fused nonlinear activation unit an \textit{analog activation} unit.
In this way, activation energy cost can be largely eliminated, as most ADCs can be replaced by energy-efficient MACAM.

Considering the precision issue, instead of supporting general nonlinear functions as in~\cite{HWA_DAC2021_Song}, we choose to implement the ReLU-$\alpha$ function.
ReLU-$\alpha$ is widely used in quantized models that have limited bit-widths to represent weight and activation.
The ReLU-$\alpha$ function with a clipping threshold $\alpha$ works as follows, 
\begin{equation}
    \small
    \label{eq:ReLUN}
    \begin{aligned}
        \widehat{\mathcal{X}} = \left\{
        \begin{aligned}
            0, &~~\mathcal{X}\in (-\infty, 0),\\
            \mathcal{X}, &~~\mathcal{X}\in[0, \alpha), \\
            \alpha, &~~\mathcal{X}\in[\alpha, +\infty), \\
        \end{aligned}
        \right.
    \end{aligned}
\end{equation}
where $\mathcal{X}$ is the pre-activation feature map and $\widehat{\mathcal{X}}$ is the final output feature map.
The clipping threshold $\alpha$ can bound the output of the ReLU function, 
therefore, a small bit-width is capable of representing the bounded value range of feature maps.

\begin{figure}
    \centering
    \includegraphics[width=0.84\columnwidth]{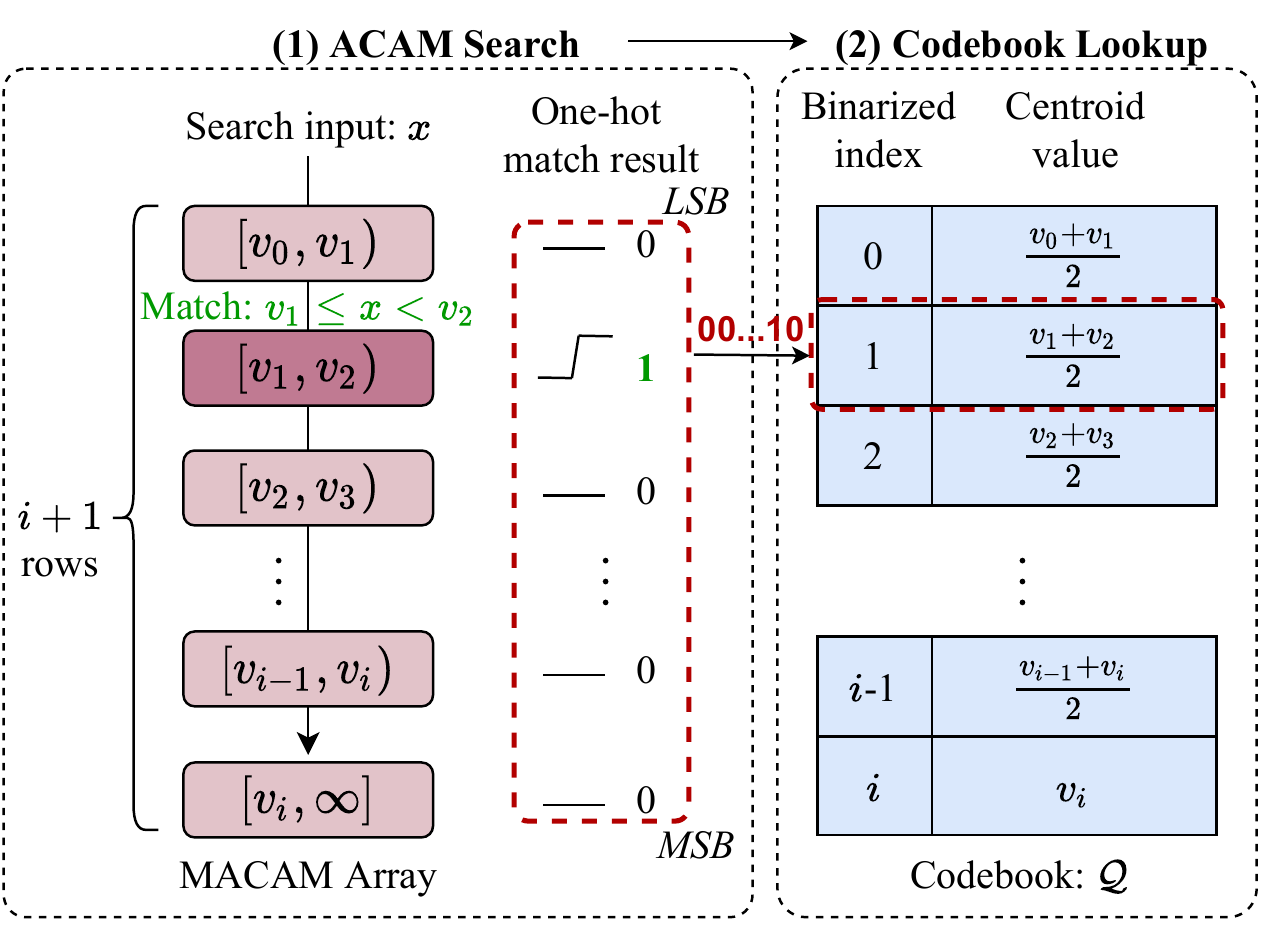}
    \caption{
    ~\small
    Illustration of implementing the positive part of ReLU-$\alpha$ and A/D conversion in a single ACAM array.
    }
    \label{fig:ReLUN_ACAM}
    \vspace{-10pt}
\end{figure}

For ReLU-$\alpha$, since the input values can be either positive or negative, we need two MACAM arrays to handle positive and negative inputs, respectively.
The negative part of ReLU-$\alpha$ is easy to implement by generating a constant digital result `0'.
Fig.~\ref{fig:ReLUN_ACAM} illustrates the implementation for the positive part of ReLU-$\alpha$ in a single MACAM array.
In the $i$-th row of the MACAM array, by tuning the resistance of the MTJ, one MACAM cell represents an acceptance search interval $[v_{i-1}, v_i)$ for a match.
All the intervals of MACAM cells in the top $i$ rows consist of a searchable range of $[v_0, v_i)$.
$i$ is decided by the number of implementable resistances of the MTJ. 
To achieve the search interval $[v_i, \infty]$, 
we encode $\infty$ similar to ~\cite{HWA_DAC2021_Song} by using a resistor with considerably larger resistance than the resistance upper bound of the MTJ.
It is sufficient to cover the possible maximum voltage output from the analog engine as it cannot be arbitrarily large. 
The match result of all match lines consists of a one-hot vector. 
After obtaining the one-hot match result, 
we then transfer it into a binarized index using a digital priority encoder such that it can be stored at a minimized memory storage cost.
The digital priority encoder can also handle corner cases when input is on the bounds of interval.
Each binarized index represents a group of inputs that falls into one specific interval.
Similar to quantization,
we set the binarized index to correspond to one shared value, in which we use the mid-point of the interval's lower bound and upper bound in our implementation.
For interval $[v_i, \infty]$, the corresponding value is set to $v_i$.
This mapping relationship can be defined as a codebook $\mathcal{Q}$ as shown in Fig.~\ref{fig:ReLUN_ACAM}.
Within this process, we actually digitize the analog input signals.
For example, in Fig.~\ref{fig:ReLUN_ACAM}, the input $x$ falls into $[v_1, v_2)$, the one-hot match result corresponds to the binarized index `1', which is treated as $\frac{v_1 + v_2}{2}$.
Since the minimum voltage $v_0$ cannot be 0, the input needs to be biased to support the search of inputs starting from 0.
In the following, for simplification of illustration, we assume $v_0 = 0$ and $v_i = c$.

\begin{figure}
    \centering
    \vspace{-5pt}
    \subfloat[]{
        \includegraphics[width=0.225\textwidth]{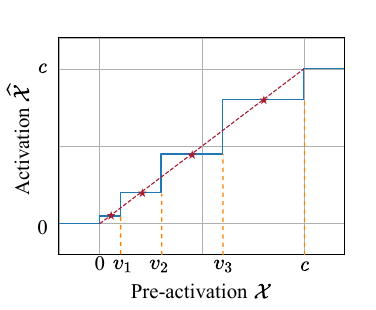}
        \label{fig:relun_curve_unreal}
    }
    \subfloat[]{
        \includegraphics[width=0.225\textwidth]{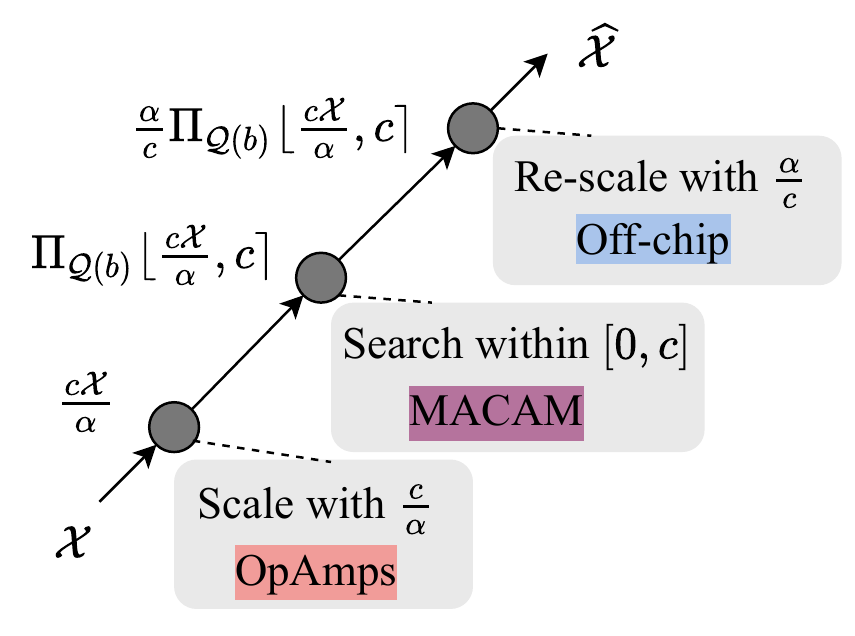}
        \label{fig:ReLUN_ACAM_flow}
    }
    \vspace{-5pt}
    \caption{~\small
    (a) Transformation behaviors of ReLU-$\alpha$ implemented with MACAM.
    (b) Implementation of arbitrary ReLU-$\alpha$ by MACAM.
    }
    \label{fig:relun_curve}
    \vspace{-10pt}
\end{figure}

Now we explain how the projection of the MACAM array can realize \emph{in-situ} nonlinearity of ReLU-$\alpha$.
Figure~\ref{fig:relun_curve_unreal} show the transformation behavior of the MACAM array.
The range $[0,c]$ is divided into four intervals, represented by the MTJ with 5 resistance levels using the measured data from ~\cite{HWA_Arixv2022_Lenoard}.
The inputs within each interval are treated as the same value after projection based on the $\mathcal{Q}$.
If we draw the red dashed line through the interval's midpoint, which is labeled by the star symbol, it implements $y=x$.
Actually, we can treat this as a nonlinear quantized version of a ReLU-$\alpha$ function.

The MACAM array can only support a fixed searchable range of $[0, c]$.
In order to implement any arbitrary ReLU-$\alpha$, we need to first scale the $\mathcal{X}$ with $\frac{c}{\alpha}$ to fit the search range, feed the scaled input into ACAM, and then scale it back to $[0, \alpha]$, as shown in Fig~\ref{fig:ReLUN_ACAM_flow}.
The first scaling operations before the MACAM array can be done using OpAmps.
We can use shared OpAmps at the output of the computation units without extra overhead~\cite{HWA_ASPDAC2020_Sun}.
The off-chip computers can do the second scaling operation.

All in all,
the behavior of MACAM-based activation can be modeled as follows,
\begin{equation}
    \small
    \label{eq:ReLUN-mtj}
    \widehat{\mathcal{X}} = \frac{\alpha}{c} \Pi_{\mathcal{Q}}\lfloor\frac{c\mathcal{X}}{\alpha}, c\rceil,
\end{equation}
where $\Pi_{\mathcal{Q}}$ indicates the discretized projection function of the ACAM within $[0, c]$ following the codebook $\mathcal{Q}$.
$\lfloor \cdot, c \rceil$ denotes clipping to the range of $[0, c]$.

\begin{figure*}
    \centering
    \includegraphics[width=0.75\textwidth]{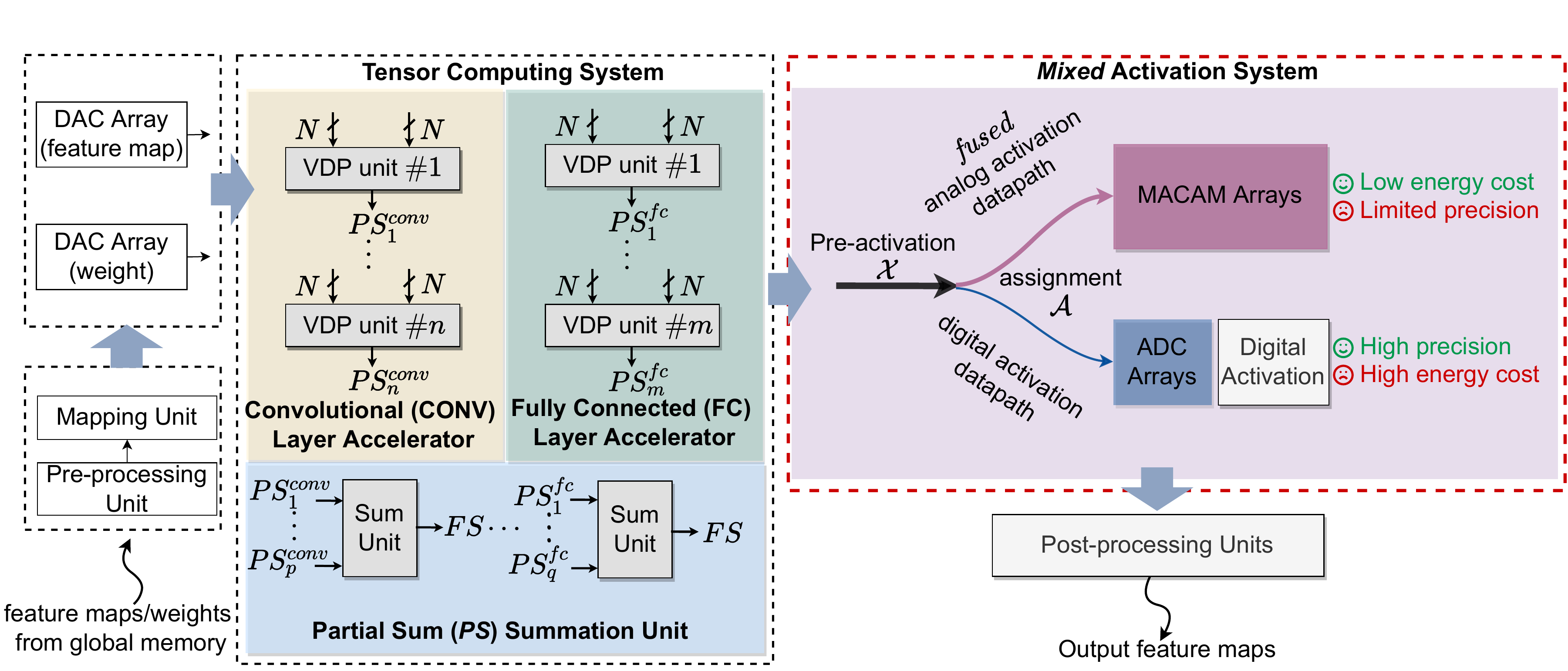}
    \caption{~\small The system architecture overview, including tensor computing system and the proposed \textit{mixed} activation system.
    }
    \label{fig:macam_system_arch}
    \vspace{-10pt}
\end{figure*}

\begin{figure}
    \centering
    \includegraphics[width=0.78\columnwidth]{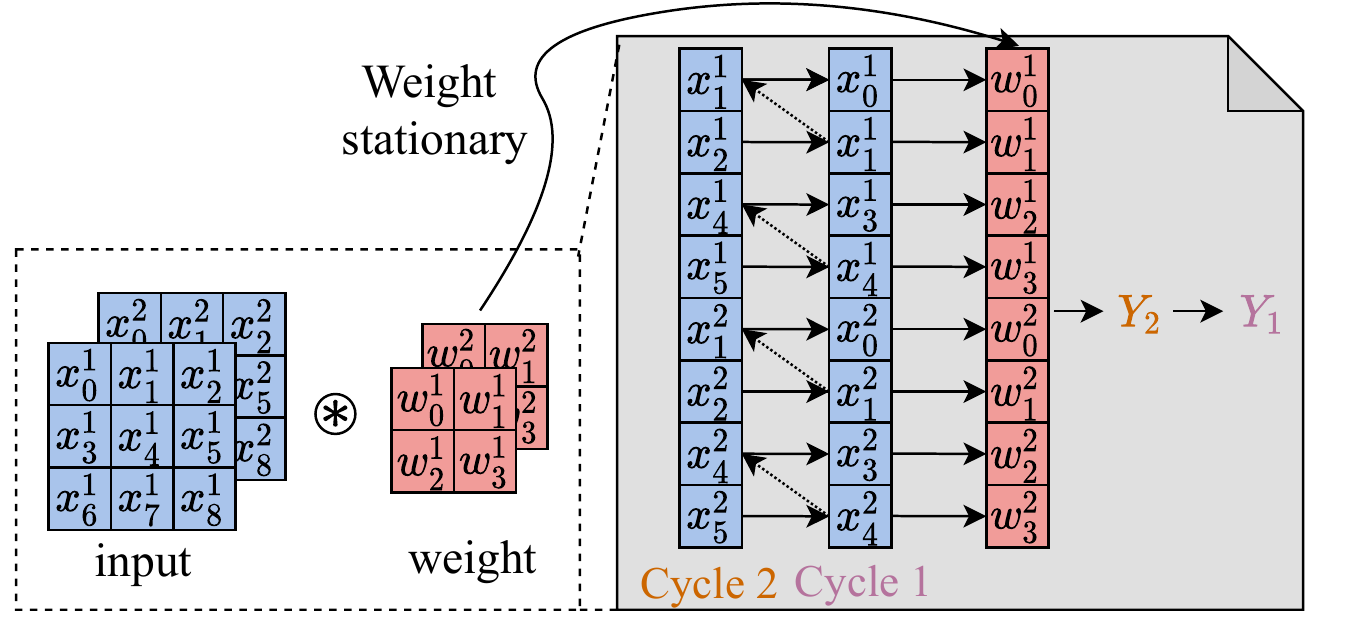}
    \caption{~\small Illustration of our adopted weight-stationary dataflow to combine convolution and activation. We show the convolution of one filter with 2 channels.
    }
    \label{fig:datflow}
    \vspace{-10pt}
\end{figure}
\section{Mixed Activation System}
\label{sec:hete_ad_sys}
\subsection{Proposed Mixed Activation System}

As the DW-MTJ can currently only demonstrate a small number of resistance levels,
the number of implementable intervals in MACAM is restricted, further limiting the representation capability of the MACAM-based analog activation unit.
Hence, we are motivated to provide a \textit{mixed} activation system, which integrates both the \textit{fused} analog activation and the traditional digital activation datapaths.
The traditional digital activation datapath first uses high-precision ADCs to do A/D conversion, then performs nonlinearity in dedicated digital activation units.
The output feature map can choose either of the two paths for nonlinear activation and A/D conversion.
With the \textit{mixed} activation system,
we can jointly utilize the low-energy analog activation datapath and the high-precision digital datapath
to balance expressiveness and energy efficiency.

Figure~\ref{fig:macam_system_arch} shows the system architecture overview with the tensor computing system and the proposed \textit{mixed} activation system.
We represent the fundamental tensor computing unit as vector-dot-product~(VDP) unit, with each unit supporting the dot-product between two length-$N$ vectors.
Multiple VDPs implement a large-size vector product due to the limited single VDP size, generating multiple partial sums (PS).
Note that nonlinear activation is not applied to any intermediate PS but to the \emph{final} computation results, thus applying our MACAM-based analog activation requires in-place partial sum accumulation in the analog domain, which is doable by partial sum summation units~\cite{HWA_MICRO2019_chou, NP_DAC2021_sunny, HWA_DAC2021_Song}.
Then,
the final results are assigned to either the analog activation or the digital activation datapath to do nonlinear activation and A/D conversion.

We elaborate a weight-stationary dataflow to meet the requirement of the overall system following~\cite{HWA_DAC2021_Song}.
Take a convolutional layer as an example, where the 2-D weights $W \in \mathbb{R}^{C_o\times (C_i k^2)}$ and 2-D output feature map $\mathcal{X} \in \mathbb{R}^{C_o\times (H^{'}W^{'})}$.
The size of vector-dot-product in its computaion is $C_ik^2$, which is distributed onto $\lceil\frac{C_ik^2}{N}\rceil$ VDPs.
This is doable since the largest vector product in modern ResNet and VGG models is only $(512\times 3 \times 3)$.
At each cycle, we obtain the convolution result of the whole dot-product, ensured by in-place partial sum summation units.
Fig.~\ref{fig:datflow} illustrates our dataflow on a simple filter with 2 input channels.
The weights can be stationary in the VDP units to continually complete the convolutions of adjacent sliding windows.
Instead of frequently fetching inputs from costly memory,
inputs can be reused in the convolutions of adjacent sliding windows.
After obtaining the whole result, it can be sent to the activation datapath to do A/D conversion and activation.

\subsection{Fully Differentiable \name Training}
Our \textit{mixed} activation system provides a more costly but higher precision digital activation datapath to compensate for the expressiveness due to analog activation datapath.
This intuitively raises the question, 
``how do we \textit{assign} the activation tasks of output feature maps onto the two different activation datapaths to balance expressiveness and energy efficiency?''

\noindent\textbf{Problem Formulation}.
Considering the energy gap between the two activation datapaths,
our target is to assign the activation task of each value of output feature map $X\in\mathcal{X}$ to either of the two activation paths with high expressiveness under a given activation energy constraint.
We define the assignment as $\mathcal{A}$.
In this way, the problem can be formulated as follows,
\begin{equation}
\label{eq:Formulation}
\small
\begin{aligned}
&\min \mathcal{L}\big(W^{*\mathcal{A}};~\mathcal{D}^{val}\big)\\
\text{s.t.} ~~&W^* = \argmin_{W} \mathcal{L} ( W^{\mathcal{A}}; ~ \mathcal{D}^{trn}), \\
&E_{act, min}\leq E_{act}(\mathcal{A}) \leq E_{act, max},\\
& \widehat{X}^{l} = \sum_{i=1}^{2} a^{l}_{X, i}f_i(X^l), ~X^{l}\in\mathcal{X}^{l},\\
&\sum_{i=1}^{2} a^{l}_{X, i} = 1, ~~a^{l}_{X, i}=\{0,1\}.
\end{aligned}    
\end{equation}
The binary selection variable $a_{X, i}$ decides that each output feature value $X \in \mathcal{X}^{l}$ in the $l$-th layer is passed through either $f_1(\cdot)$ (MACAM-based analog activation datapath) or $f_2(\cdot)$~(digital activation path). The selection variables consist of the assignment $\mathcal{A}$, which is our primary search target with a activation energy cost $E_{act}(\mathcal{A})$ satisfying its constraints $[E_{act, min}, E_{act, max}]$.

\noindent\textbf{Search Space Specification}.~
Considering the on-chip routing issue and computation regularity,
it is not realistic to tediously assign individual output feature map value to different activation units.
Instead, in this work, we propose to assign the workloads in the \textit{filter} level.
Concretely, take a convolutional layer $l$ as an example.
Suppose it contains $C_o$ filters and each filter has $C_{i}$ input channels and kernel size $k$, where its 2-D input $\mathcal{X}^{l-1}\in \mathbb{R}^{C_i\times (H \times W)}$ and 2-D output feature $\mathcal{X}^{l}\in \mathbb{R}^{C_o\times (H^{'} \times W^{'})}$.
The entire $b$-th output channel of the output feature map $\mathcal{X}^{l}$ is passed through either the analog activation path $f_1(\cdot)$ or the electrical activation path $f_2(\cdot)$, decided by a binary selection variable $a_{b, i}^{l}$.
The behavior is given as follows,
\begin{equation}
    \small
    \label{eq:select_beh}
    \widehat{\mathcal{X}}^{l}_{b} = \sum_{i=1}^{2} a_{b,i}^{l}f_i(\mathcal{X}^{l}_{b}), \ a_{b, i}^{l}\in \{0,1\}, \ \sum_i a_{b, i}^{l} = 1.
\end{equation}

In this way, our search space is defined as the assignment $\mathcal{A}$ that assigns each channel of output feature maps to either the analog or the electrical activation path.
For layer $l$, the number of total combinations can be $2^{C_o}$.
Thus,
the total search space for a model with $L$ layers is extremely large, which is $\mathcal{O}(\prod_{l}^{L}2^{C_o^{l}})$.

\begin{figure}
    \centering
    \includegraphics[width=0.95\columnwidth]{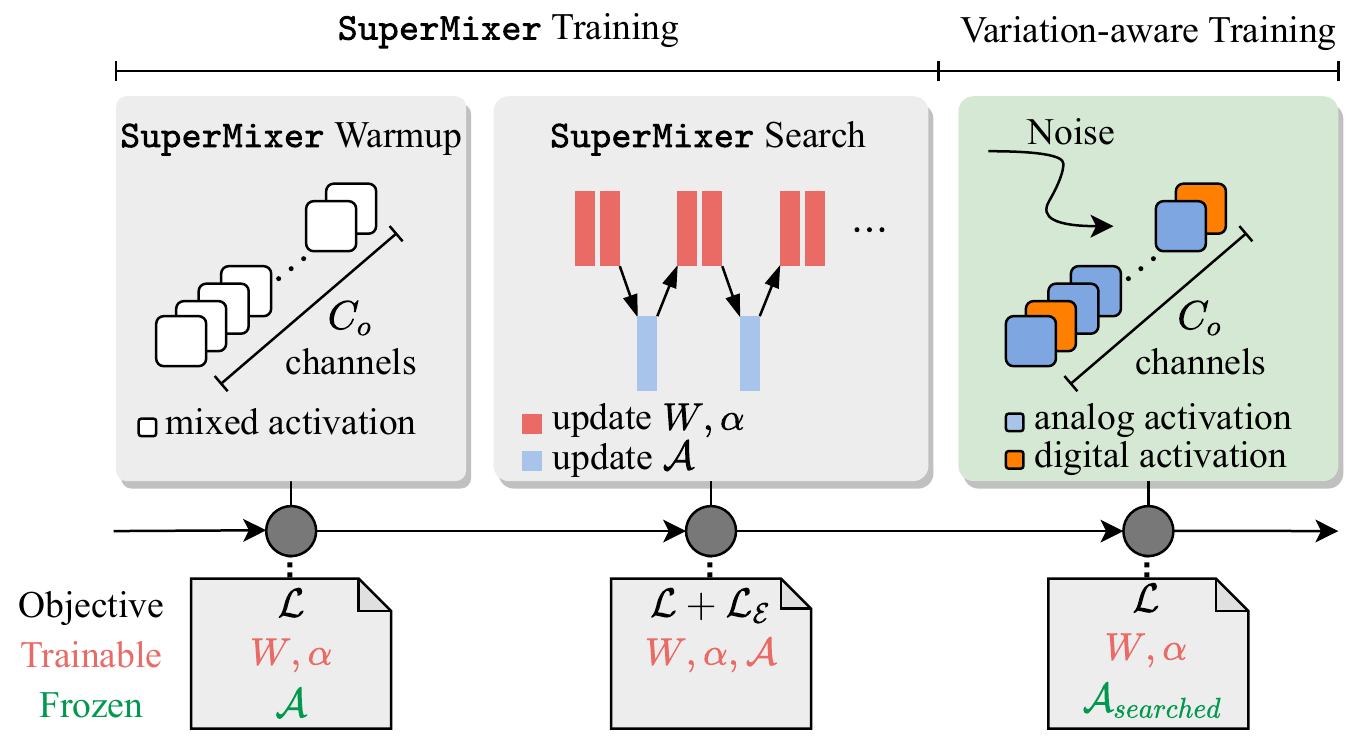}
    \caption{
    ~\small The proposed \name training flow and variation-aware training with injected noises.
    }
    \label{fig:supermixer_flow}
    \vspace{-10pt}
\end{figure}

\noindent\textbf{Fully Differentiable \name Training}.~
Considering the enormous search space of $\mathcal{A}$ and its discreteness,
we propose a differentiable \name training flow as shown in Fig.~\ref{fig:supermixer_flow}.

In the \name training, we need to optimize weight $W$,  
clipping threshold $\alpha$ in ReLU-$\alpha$, 
and assignment $\mathcal{A}$.
It is highly ill-conditioned to jointly optimize all those continuous and discrete variables.
Aware of this, in this work, we divide our \name flow into two phases.
The first \name \textit{Warmup} phase aims at stabilizing our following search stage, 
where only $W$ and $\alpha$ are optimized to obtain a good initial point.
The second \name \textit{Search} phase aims at searching for the assignment $\mathcal{A}$ to boost the expressiveness.
It optimizes (1) $W$, $\alpha$ and (2) $\mathcal{A}$ alternatively to avoid prohibitive co-optimization difficulty.
We periodically enter the optimization of (1) and (2) with a ratio of 2:1.
Moreover,
during the \name \textit{Search} phase, we need to satisfy the activation energy cost constraint,
thus, an energy cost penalty $\mathcal{L_E}$ is added besides the original loss $\mathcal{L}$.
After \name training, we fix the searched optimal assignment $\mathcal{A}_{searched}$ and conduct variation-aware training to improve the model's robustness regarding on-chip variations, e.g., the resistance variation of memristors in MACAM.

Now we explain how to optimize assignment $\mathcal{A}$ in a differentiable way.
As shown in Eq.~\eqref{eq:select_beh}, the selection variable $a_{b,i}^l$ is a binary variable.
Instead of searching $\mathcal{A}$ in such a discrete space, we relax the optimization problem by constructing a stochastic mixed activation unit.
During the inference,
the activation of the $b$-th channel of feature in layer $l$, $\mathcal{X}_b^{l}$, is using either the analog activation path $f_1(\cdot)$ or the digital activation path $f_2(\cdot)$ with the sampling probability of
\begin{equation}
    \small
    \label{eq:select_prob}
    P_{\theta_b^{l}}(f_b=f_i) = \frac{e^{\theta_{b,i}^{l}}}{\sum_{i}e^{\theta_{b,i}^{l}}}.
\end{equation}
Equivalently, the output of the stochastic mixed activation, $\widehat{\mathcal{X}}_{b}^{l}$, can be expressed as,
\begin{equation}
    \small
    \label{eq:select_beh2}
    \widehat{\mathcal{X}}^{l}_{b} = \sum_{i=1}^{2} a_{b,i}^{l}f_i(\mathcal{X}^{l}_{b}),
\end{equation}
where $a_{b,i}$ is a random variable in $\{0, 1\}$ and is evaluated based on the sampling probability in~\eqref{eq:select_prob}.
Therefore, through parameterizing the probability distribution of activation unit choices by the sampling coefficient $\theta_b^l$, 
we can relax the problem as the optimization of probability of $P_{\theta}$.
However, we cannot propagate the gradient back through the discrete variable $a_{b,i}$ to $\theta_{b,i}$.
To sidestep this issue, Gumbel-Softmax (GS) trick is adopted to relax $a_{b,i}$ to be a continuous varible as follows,
\begin{equation}
    \small
    a_{b,i}^{l} = \text{GumbelSoftmax}(\theta_{b,i}^{l}|\mathbf{\theta_{b}^{l}}) =\frac{e^{(\theta_{b,i}^{l}+g_{b,i}^{l})/\tau}}{\sum_{i}e^{(\theta_{b,i}^{l}+g_{b,i}^{l})/\tau}}.
\end{equation}
$g_{l,i}^l$ follows the Gumbel distribution $\text{Gumbel}(0,1)$ as a Gumbel noise.
A temperature parameter $\tau$ is used to control the GS function.
As $\tau$ is close to $0$,
the Gumbel-Softmax function approximates categorical samples based on~\eqref{eq:select_prob}.
A larger $\tau$ introduces randomness to encourage exploitation of the assignment $\mathcal{A}$. 
Therefore, during the \name \textit{Search} phase, we gradually decay $\tau$ such that \name can automatically exploit the search space and learn the optimal assignment to augment $\mathcal{A}$ the models' expressiveness.

\noindent\textbf{Activation Energy-Constrained Optimization}.~
The assignment $\mathcal{A}$ directly impacts the activation energy cost, as it defines the mixed way of using low-precision analog activation units and high-precision electrical activation units.
Fully using analog activation datapath results in an ADC-free system to convert computation results to the digital domain, 
while fully using electrical activation datapath raised a serious A/D conversion cost.
With the assignment $\mathcal{A}$, we can get the activation energy cost $E(\mathcal{A})$ as follows,
\begin{equation}
    \label{eq:act_energy}
    \small
    \begin{aligned}
        E_{act}({\mathcal{A}}) &= \#act_{anlg} \cdot E_{anlg} + \#act_{digi} \cdot E_{digi}\\
        &=\sum_{l}^{L}\sum_{b}^{C_o}(a_{b, i}^l \cdot E_{anlg} + a_{b, 2}^l \cdot E_{digi}) \cdot H^{'} W^{'}.
    \end{aligned}
\end{equation}
$\#act_{anlg}$ and $\#act_{digi}$ denote the number of output feature maps being passed through the analog (\emph{anlg}) activation datapath and the digital (\emph{digi}) activation datapath, respectively. 
The $E_{anlg}$ and $E_{digi}$ represent the energy cost of two datapaths.
The former indicates the MACAM array's cost, and the latter contains ADC and digital activation unit costs.
As the ADC cost is far larger~\cite{HWA_ISCA2016_Shafiee} and the digital activation unit cost is dependent on the digital part's frequency,
we use ADC cost to represent $E_{digi}$ during the search phase.

To honor the energy cost constraint $[E_{act, min}, E_{act, max}]$,
we add a \textit{probabilistic activation energy cost penalty} term $\mathcal{L_E}$,
\begin{equation}
    \small
    \label{eq:energy_loss}
    \begin{aligned}
        &\mathcal{L_E} = \left\{
        \begin{aligned}
            \beta\left(E_{act}(\mathcal{A})/(1-\gamma)E_{max}\right), &~~E_{act}(\mathcal{A}) > (1-\gamma)E_{max},\\
           -\beta\left(E_{act}(\mathcal{A})/(1+\gamma)E_{min}\right), &~~E_{act}(\mathcal{A}) < (1+\gamma)E_{min},\\
            0, &~~\text{otherwise}.\\
        \end{aligned}
        \right. \\
    \end{aligned}
\end{equation}
We set the margin $\gamma$ to $5\%$ to tighten the constraint.
Note that $a_{b,i}^l$ in Eq.~\eqref{eq:act_energy} is differentiable guaranteed by the GS trick.

\noindent\textbf{Precision-Adaptive $\alpha$ Optimization}.~
Implementing ReLU-$\alpha$ can ease the issue of insufficient precision of MACAM.
However, it is not enough to address this since the SOTA MTJ can only provide five stable resistance levels, as mentioned before.
The precision of ReLU-$\alpha$ implemented by ACAM is only around $2$-bit.
Thus, a solution is in great need to further tolerate the precision issue.

Instead of choosing a fixed $\alpha$ in Eq.~\eqref{eq:ReLUN}, inspired
by PACT~\cite{NN_arxiv2018_Choi}, we propose to adopt a \emph{learnable} $\alpha$ with an \emph{enhanced training recipe} to \textit{accommodate the low resolution of MACAM for accuracy boost.}
This can be done by configuring the gain of the OpAmps~\cite{HWA_ASPDAC2020_Sun}.
In~\cite{NN_arxiv2018_Choi},
during the learning process of $\alpha$,
the gradient to $\alpha$ is computed by $\frac{\partial \widehat{\mathcal{X}}}{\partial \alpha} = \texttt{Sign}(\mathcal{X})$ when $\mathcal{X} \geq \alpha$.
Activation $\mathcal{X}$ that is smaller than $\alpha$ cannot contribute to the gradient, resulting in inaccurate gradient estimation to $\alpha$.
Instead of updating $\alpha$ in the same way of~\cite{NN_arxiv2018_Choi}, we re-formulate the gradient to $\alpha$ based on Eq.\eqref{eq:ReLUN-mtj} as follows,
\begin{equation}
    \small
    \label{eq:ReLUN_grad_alpha_mtj}
    \begin{aligned}
        \frac{\partial \widehat{\mathcal{X}}}{\partial \alpha} &= \frac{\partial \alpha}{\partial \alpha} \cdot \frac{1}{c} \Pi_{\mathcal{Q}(b)}\lfloor\frac{c\mathcal{X}}{\alpha}, c\rceil + \frac{\partial \Pi_{\mathcal{Q}(b)}\lfloor\frac{c\mathcal{X}}{\alpha}, c\rceil}{\partial \lfloor\frac{c\mathcal{X}}{\alpha}, c\rceil}\frac{\partial \lfloor\frac{c\mathcal{X}}{\alpha}, c\rceil}{\partial \alpha} \cdot \frac{\alpha}{c}\\
        &=\left\{
        \begin{aligned}
            0, &~~\mathcal{X}\in (-\infty, 0),\\
            1 \cdot \frac{1}{c} \Pi_{\mathcal{Q}(b)}\lfloor\frac{c\mathcal{X}}{\alpha}, c\rceil + 1\cdot \frac{-c\mathcal{X}}{\alpha^2} \cdot \frac{\alpha}{c}, &~~\mathcal{X}\in[0, \alpha), \\
            1 \cdot \frac{1}{c}\cdot \texttt{Sign}(\frac{c\mathcal{X}}{\alpha})\cdot c + \frac{\alpha}{c} \cdot 0, &~~\mathcal{X}\in[\alpha, +\infty), \\
        \end{aligned}
        \right. \\
        &=\left\{
        \begin{aligned}
            0, &~~\mathcal{X}\in (-\infty, 0),\\
            \frac{1}{c} \Pi_{\mathcal{Q}(b)}\lfloor\frac{c\mathcal{X}}{\alpha}, c\rceil - \frac{\mathcal{X}}{\alpha}, &~~\mathcal{X}\in[0, \alpha) ,\\
            1, &~~\mathcal{X}\in[\alpha, +\infty), \\
        \end{aligned}
        \right.\\
    \end{aligned}
\end{equation}
where our scaling operation works as a reparameterization trick to preserve the gradient contribution from $\mathcal{X} \in [0, \alpha)$.
Thus, it can correct the inaccurate gradient estimation to $\alpha$, which is proved to get a significantly larger accuracy boost than~\cite{NN_arxiv2018_Choi} in our experiments.
\section{Case study: Photonic Accelerator}
\label{sec:case_study}

\begin{figure}
    \centering
    \includegraphics[width=0.88\columnwidth]{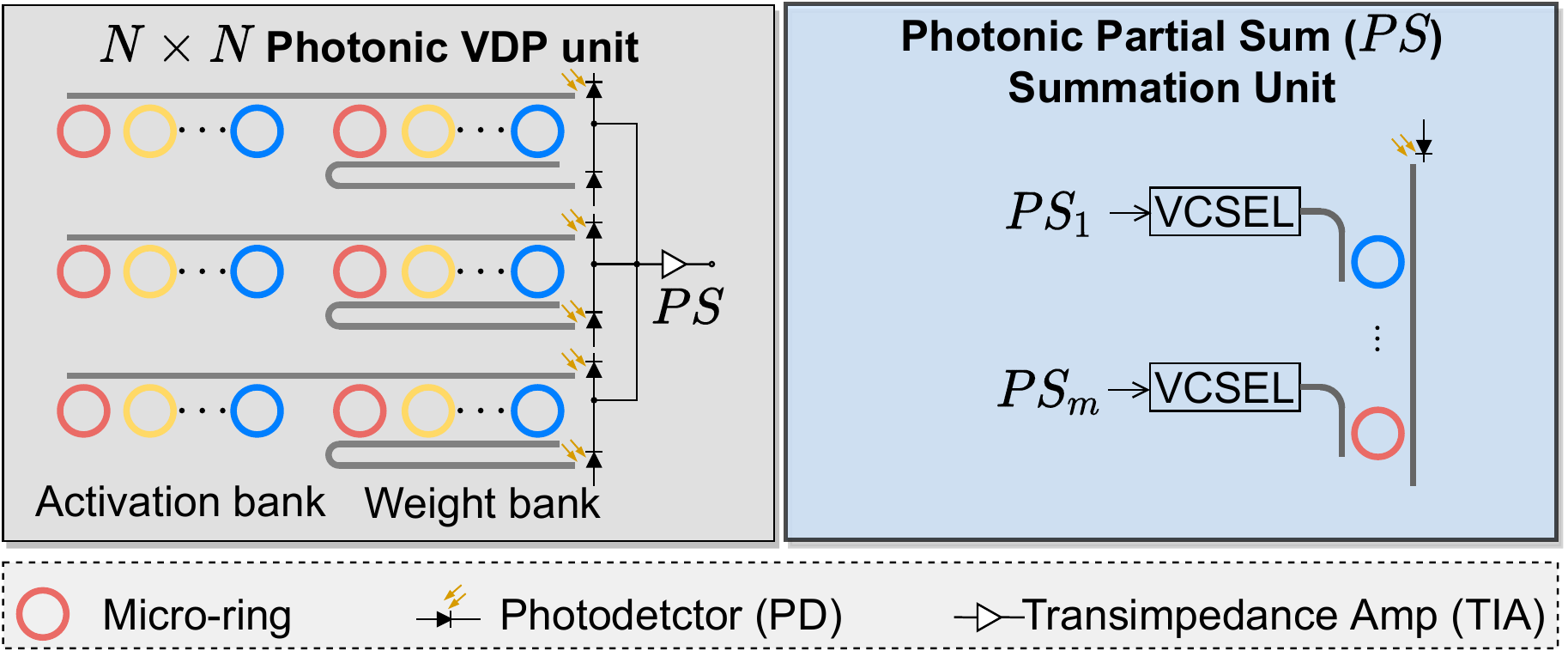}
    \caption{~\small The adopted photonic vector-dot-product unit~\cite{NP_SciRep2017_Tait,NP_DAC2021_sunny} and photonic partial sum summation unit~\cite{NP_DAC2021_sunny}.
    }
    \label{fig:photonic_case_study}
    \vspace{-15pt}
\end{figure}

We demonstrate our design principle on a photonic accelerator as a case study.
We focus on a SOTA incoherent photonic design based on micro-ring (MR) resonators~\cite{NP_SciRep2017_Tait, NP_DAC2021_sunny}.
Other accelerators can also benefit from our methods.
For example,
we can simply replace the ACAM design in a RRAM-based accelerator~\cite{HWA_DAC2021_Song} with ours but own better tolerance to signal noise and the precision of ACAM.

Fig.~\ref{fig:photonic_case_study} shows the adopted photonic vector-dot-product (VDP) engine~\cite{NP_SciRep2017_Tait, NP_DAC2021_sunny} and photonic in-place partial sum (PS) summation unit~\cite{NP_DAC2021_sunny} for our system in Fig. \ref{fig:macam_system_arch}.
The computation of the convolutional~(CONV) layer and fully-connected~(FC) layer are unfolded to matrix multiplication, where each vector dot product is implemented by the vector dot product~(VDP) units based on micro-ring~(MR) resonators.
To support two length-$N$ vector multiplication within each VDP unit, the $N$ size vectors are decomposed into small-sized vector chunks.
Each small-sized vector dot-product is performed using MRs in each arm of the VDP unit.
In this way, a large $N\times N$ vector dot product can be achieved in one unit, e.g., $100\times100$ in \cite{NP_DAC2021_sunny}. 
Across VDP units, 
photonic partial summation units are used to accumulate the partial sums from multiple VDPs.
The partial sums from multiple VDPs are converted from the analog domain to the photonic domain by VCSELs, multiplexed into one single waveguide, and summed via another photodetector.

\noindent\textbf{Energy Modeling of the A/D Conversion and Activation Cost}.
Here, we model the energy cost of the A/D conversion and activation, $E_{A/D+act}$ based on \texttt{O-HAS}~\cite{NP_ICCAD2021_Li}.
Our mixed activation system provides two datapaths to do nonlinear activation and A/D conversion, with an extra photonic summation unit overhead.
Consider one convolutional layer with $C_o$ filters and output feature map $\mathcal{X}^{l}\in \mathbb{R}^{C_o\times (H^{'} \times W^{'})}$.
Each filter has $C_i$ input channels and a kernel size of $k$.
The energy model of $E_{A/D+act}$ can be modeled as follows,
\begin{equation}
    \small
    \label{eq:ad_energy_ours}
    \begin{aligned}
        E_{A/D+act} &= E_{act}(\mathcal{A}) + E_{sum}\\
        &=(C_{o, anlg}E_{anlg} + C_{o, digi}(E_{digi, adc} + E_{digi, act}))H^{'}W^{'} \\
        &+ (E_{VCSEL})C_oH^{'}W^{'}\Big\lceil\frac{C_ik^2}{N}\Big\rceil + C_oH^{'}W^{'}E_{PD}.
    \end{aligned}
\end{equation} 
$E_{anlg}$ denotes the energy of the analog activation datapath, $E_{digi, adc}$ and $E_{digi, act}$ denote the A/D conversion and activation energy of the digital activation datapth.
$E_{VCSEL}$ and $E_{PD}$ denote the energy for VCSEL and photodetector in the photonic PS summation unit.

In conventional implementation, the A/D conversion is done right after obtaining partial sum and the digital activation datapath is fully used. The energy cost can be modeled as,
\begin{equation}
    \small
    \label{eq:ad_energy_normal}
    \begin{aligned}
        E_{A/D+act} &= E_{act} + E_{A/D} \\
        &=C_oH^{'}W^{'}(E_{digi, act}) \\
        &+ (E_{ADC}+E_{S+A}+E_{PD})C_oH^{'}W^{'}\Big\lceil\frac{C_ik^2}{N}\Big\rceil.
    \end{aligned}
\end{equation}
\section{Experimental Results}
\label{sec:ExperimentalResults}
\subsection{Experiment Setup}
\label{sec:ExpSetup}
\noindent\textbf{Models and Datasets}.~
We evaluate our methods on two modern CNNs (VGG13 and ResNet18) and CIFAR100~\cite{NN_cifar2009}, with 6-bit weight precision.
In VGG13,  we replace the last three FC layers with one to avoid over-fitting.
In ResNet18, we move the residual path after activation such that no extra addition is needed before activation.

\noindent\textbf{Training Settings}.~
In \name training flow, we train for 90 epochs using an SGD optimizer with an initial learning rate~(lr) of 0.02, a momentum of 0.9, and a cosine lr scheduler.
The Gumbel-softmax temperature $\tau$ exponentially decreases from $5$ to $0.5$.
$\gamma$ is set to 0.6 for the activation energy cost penalty.
The initial $\alpha$ of the adaptive ReLU-$\alpha$ is set to 8.
We set 10 epochs for \name  \textit{Warmup} phase and 80 epochs for \textit{search} phase.
We sample an assignment $\mathcal{A}$ from the learned distribution $P_{\theta}$, then enter the variation-aware training.
We train all models for 200 epochs during variation-aware retraining using an SGD optimizer with an initial learning rate~(lr) of 0.02, a momentum of 0.9, and a cosine lr scheduler.

\begin{table}[]
\centering
\caption{~\small 6-bit ADC configurations considered in this paper.}
\vspace{-5pt}
\label{tab:adc}
\resizebox{0.65\columnwidth}{!}{%
\begin{tabular}{c|cc}
\toprule
              & ADC-1~\cite{NP_ISSCC2016_Choo} & ADC-2~\cite{NP_Alphacore2022} \\ \midrule
Bit-width           & 6     & 6     \\
Sampling rate~($\text{GS/s}$) & 1     & 6     \\
Power~($\text{mW}$)         & 1.26  & 14   \\ 
Latency~($\text{ns}$)       & 8      & 1.33     \\
\bottomrule
\end{tabular}%
}
\vspace{-10pt}
\end{table}

\noindent\textbf{MACAM and ADC Designs}.~
We consider two MACAM designs with two MTJ devices in this paper.
MACAM-1 uses the DW-MTJ with 5 resistance levels with around 2-bit precision. MACAM-2 uses the DW-MTJ with 3 resistance levels with around 1-bit precision.
We choose two 6-bit ADC designs shown in Table.~\ref{tab:adc}.

\noindent\textbf{Noise Injection}.~
We set the variation of MRs following a Gaussian distribution $\mathcal{N}(0,0.05^2)$.
For MACAM, we set the MTJ resistance device-to-device variation as $\mathcal{N}(0,0.128^2)$~\cite{HWA_APL2021_Liu}, run Monte Carlo simulation 10$^4$ times to capture the noisy distribution of MACAM's intervals, and equivalently add it to the input following~\cite{HWA_DAC2021_Song}.

\noindent\textbf{A/D and Activation Energy Simulation}.~
The performance and power dissipation of MACAM are evaluated with Cadence ADE and spectre simulations.
We estimate the A/D conversion and activation energy by Eq.~\eqref{eq:ad_energy_ours} based on \texttt{O-HAS}~\cite{NP_ICCAD2021_Li}.
The size of VDP is set as 128. 
Since the final FC layer is the output layer without activation, 
for a fair comparison,
we don't include its energy consumption.

\subsection{Main Results}
\label{sec:MainResults}

\begin{table}[]
\centering
\caption{~\small Compare different ReLU variants on CIFAR100.
$2+$ and $1+$ represents the precision of ACAM-1 and ACAM-2, respectively.
}
\vspace{-5pt}
\label{tab:CompareReLU}
\resizebox{0.88\columnwidth}{!}{%
\begin{tabular}{c|cc|c|c}
\toprule
Model                    & Weight bit                & Act. bit                   & ReLU                                                                & Accuracy (\%)                                           \\ \midrule
                         & 32                        & 32                         & ReLU                                                                & 74.76                                                   \\ \cmidrule{2-5}
                         & 6                         & 6                          & ReLU6                                                               & 71.31                                                   \\
                         & 6                         & 6                          & ReLU-$\alpha$-PACT~\cite{NN_arxiv2018_Choi} & 73.80                                                   \\
                         & \cellcolor[HTML]{C0C0C0}6 & \cellcolor[HTML]{C0C0C0}6  & \cellcolor[HTML]{C0C0C0}ReLU-$\alpha$-Ours                          & \cellcolor[HTML]{C0C0C0}\textbf{73.91} \\ \cmidrule{2-5}
                         & 6                         & 2+                         & ReLU6                                                               & 68.17                                                   \\
                         & 6                         & 2+                         & ReLU-$\alpha$-PACT~\cite{NN_arxiv2018_Choi} & 70.96                                                   \\
                         & \cellcolor[HTML]{C0C0C0}6 & \cellcolor[HTML]{C0C0C0}2+ & \cellcolor[HTML]{C0C0C0}ReLU-$\alpha$-Ours                          & \cellcolor[HTML]{C0C0C0}\textbf{72.52} \\ \cmidrule{2-5}
                         & 6                         & 1+                         & ReLU2                                                               & 54.98                                                   \\
                         & 6                         & 1+                         & ReLU-$\alpha$-PACT~\cite{NN_arxiv2018_Choi} & 67.11                                                   \\
\multirow{-10}{*}{VGG13} & \cellcolor[HTML]{C0C0C0}6 & \cellcolor[HTML]{C0C0C0}1+ & \cellcolor[HTML]{C0C0C0}ReLU-$\alpha$-Ours                          & \cellcolor[HTML]{C0C0C0}\textbf{70.84} \\ \bottomrule
\end{tabular}
}
\vspace{-10pt}
\end{table}
\noindent\textbf{Evaluation of Our Precision-adaptive ReLU-$\alpha$}.~
In Table~\ref{tab:CompareReLU}, we validate the expressiveness of our proposed precision-adaptive ReLU-$\alpha$ on VGG13.
We compare it with the adaptive ReLU-$\alpha$ in PACT~\cite{NN_arxiv2018_Choi} and the ReLU-$\alpha$ with a fixed $\alpha$.
For the ReLU-$\alpha$ with fixed $\alpha$,
the commonly used ReLU6 is adopted, while ReLU2 is adopted in the extremely low activation bit-width.
Under $6$-bit weight bit-width,
our implementation achieves the highest accuracy, especially on low activation bit-width cases.
This attributes to our proposed precision-adaptive $\alpha$ optimization scheme, which learns the $\alpha$ to accommodate the low resolution.
Compared to PACT~\cite{NN_arxiv2018_Choi}, our enhanced training recipe can correct its inaccurate gradient, resulting in a better accuracy boost.
Hence,
our adaptive ReLU-$\alpha$ can address the accuracy drop issue in~\cite{HWA_DAC2021_Song} due to the low activation bit-width.
It is essential to boost the expressiveness of the analog activation unit so as to achieve competitive model accuracy.

\begin{table*}[]
\centering
\caption{~\small Test Accuracy of searched VGG13 on different ADC and MACAM designs, where the model is searched on CIFAR100.
All activation energy cost is normalized by the activation energy of fully using digital activation datapath. }
\vspace{-5pt}
\label{tab:CompareSeachVGG}
\resizebox{0.97\textwidth}{!}{%
\begin{tabular}{c|c|c|c|cc|cccc}
\toprule
Model & MACAM design &ADC design              & Metrics          & Fully digital  & Fully analog & Searched-$\mathcal{A}_{0}$       & Searched-$\mathcal{A}_{1}$         & Searched-$\mathcal{A}_{2}$      & Searched-$\mathcal{A}_{3}$     \\ \midrule
\multirow{13}{*}{VGG13}& \multirow{6}{*}{MACAM-1}& \multirow{3}{*}{ADC-1}      & {[}$E_{act, min}$, $E_{act, max}${]} & -        & -      & {[}0.05, 0.15{]}       & {[}0.15, 0.25{]} & {[}0.25, 0.35{]} & {[}0.35, 0.45{]} \\
&    &                       & Activation energy $E_{act}(\mathcal{A})$          & 1         & 0.00036 & 0.11     & 0.19      & 0.28      & 0.39      \\
&    &                       & Accuracy (\%)                    & 73.91     & 72.52   & 72.94    & 72.73     & 73.09     & \textbf{73.30}     \\ \cmidrule{3-10} 
&   &\multirow{3}{*}{ADC-2}     & {[}$E_{act, min}$, $E_{act, max}${]} & -        & -     & {[}0.09, 0.14{]}   & {[}0.14, 0.19{]} & {[}0.19, 0.24{]} & {[}0.24, 0.30{]} \\
&   &                           & Activation energy $E_{act}(\mathcal{A})$       & 0.54      & 0.00036   & 0.10     & 0.15      & 0.21       & 0.24    \\
&   &                           & Accuracy (\%)                 & 73.91     & 72.52     & 72.58     & 72.86     & 73.24     & \textbf{73.54}      \\ \cmidrule{2-10} 
& \multirow{6}{*}{MACAM-2}& \multirow{3}{*}{ADC-1}      & {[}$E_{act,min}$, $E_{act,max}${]} & -        & -      & {[}0.05, 0.15{]}       & {[}0.15, 0.25{]} & {[}0.25, 0.35{]} & {[}0.35, 0.45{]} \\
&    &                       & Activation energy $E_{act}(\mathcal{A})$          & 1         & 0.00022   & 0.12     & 0.19      & 0.27      & 0.39    \\
&    &                       & Accuracy (\%)                    & 73.91     &70.84      & 71.50     & 72.07     & 72.13     & \textbf{72.50}      \\ \cmidrule{3-10} 
& &\multirow{3}{*}{ADC-2}     & {[}$E_{act,min}$, $E_{act,max}${]} & -        & -     & {[}0.09, 0.14{]}   & {[}0.14, 0.19{]} & {[}0.19, 0.24{]} & {[}0.24, 0.30{]} \\
&           &            & Activation energy $E_{act}(\mathcal{A})$      & 0.54      & 0.00022       &0.10     & 0.16      & 0.21       & 0.26    \\
&           &            & Accuracy (\%)                & 73.91     & 70.84         & 71.50       & 71.76         & 72.42         & \textbf{73.02}      \\ \bottomrule
\end{tabular}%
}
\vspace{-5pt}
\end{table*}

\begin{table*}[]
\centering
\caption{~\small Test Accuracy of searched ResNet18 on CIFAR100 with MACAM-1 and two different ADC designs.}
\vspace{-5pt}
\label{tab:CompareSeachRes}
\resizebox{0.97\textwidth}{!}{%
\begin{tabular}{c|c|c|c|cc|cccc}
\toprule
Model & MACAM design &ADC design              & Metrics          & Fully digital  & Fully analog & Searched-$\mathcal{A}_0$       & Searched-$\mathcal{A}_1$        & Searched-$\mathcal{A}_2$     & Searched-$\mathcal{A}_3$    \\ \midrule
\multirow{6}{*}{ResNet18}& \multirow{6}{*}{MACAM-1}& \multirow{3}{*}{ADC-1}      & {[}$E_{act,min}$, $E_{act,max}${]} & -        & -      & {[}0.05, 0.15{]}       & {[}0.15, 0.25{]} & {[}0.25, 0.35{]} & {[}0.35, 0.45{]} \\
&    &                       & Activation energy $E_{act}(\mathcal{A})$          & 1         & 0.00036 & 0.12     & 0.20      & 0.29      & 0.41      \\
&    &                       & Accuracy (\%)                    & 77.63     & 76.41   & 77.01    & 77.18     & 77.35     & \textbf{77.47}     \\ \cmidrule{3-10} 
&   &\multirow{3}{*}{ADC-2}     & {[}$E_{act,min}$, $E_{act,max}${]} & -        & -     & {[}0.09, 0.14{]}   & {[}0.14, 0.19{]} & {[}0.19, 0.24{]} & {[}0.24, 0.30{]} \\
&   &                           & Activation energy $E_{act}(\mathcal{A})$       & 0.54      & 0.00036   & 0.11     & 0.16      & 0.21       & 0.27    \\
&   &                           & Accuracy (\%)                 & 77.63     & 76.41     & 76.85    & 77.06     & 77.14     & \textbf{77.40}      \\ \bottomrule
\end{tabular}%
}
\vspace{-10pt}
\end{table*}

\noindent\textbf{Evaluation of Our \name}.~
We search the assignment $\mathcal{A}$ with the proposed \name flow on different MACAM designs, ADC designs, and activation energy constraints.
We denote searched assignments as searched $\mathcal{A}_{0}$ to searched $\mathcal{A}_{3}$.
Table~\ref{tab:CompareSeachVGG} and ~\ref{tab:CompareSeachRes} show the test accuracy of searched assignments on VGG13 and ResNet18.
Our searched assignment series show improved expressiveness in terms of accuracy with constrained activation energy cost.
Especially, given a large activation energy cost budget on MACAM-1,
our \name can find an assignment with comparable accuracy to the case fully using high-precision digital activation.
In conclusion, our \name flow successfully utilizes the provided \textit{mixed} activation system to boost the expressiveness.

\noindent\textbf{Energy Saving on AD Conversion and Activation.}
We simulate the energy cost of A/D conversion and activation cost on VGG13.
Configurations of ADC-1 and ACAM-1 are adopted.
The conventional implementation of using ADCs for A/D conversion of partial sums and using digital activation datapath consumes 35.7 $\mu\text{J}$.
With photonic summation units to eliminate A/D conversion needs of partial sums, 
fully using digital activation datapath consumes 17.24 $\mu\text{J}$ with a 51.7\% reduction, and fully using analog activation datapath uses 12.44 $\mu\text{J}$ with a 65.2\% reduction.
Our \name enables the mixed use of electrical and analog activation with a trade-off between energy cost and expressiveness, where
the searched-$\mathcal{A}_3$ consumes 14.2$\mu\text{J}$ with 60.2\% reduction but comparable accuracy.

\begin{figure}
    \centering
    \vspace{-10pt}
    \subfloat[]{\includegraphics[width=0.22\textwidth]{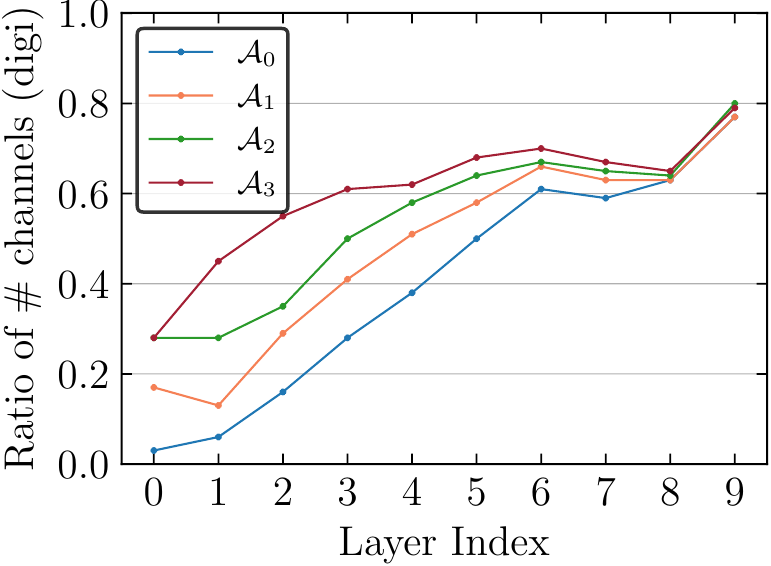}
    \label{fig:VGG13_adcR}
    }
    \subfloat[]{\includegraphics[width=0.22\textwidth]{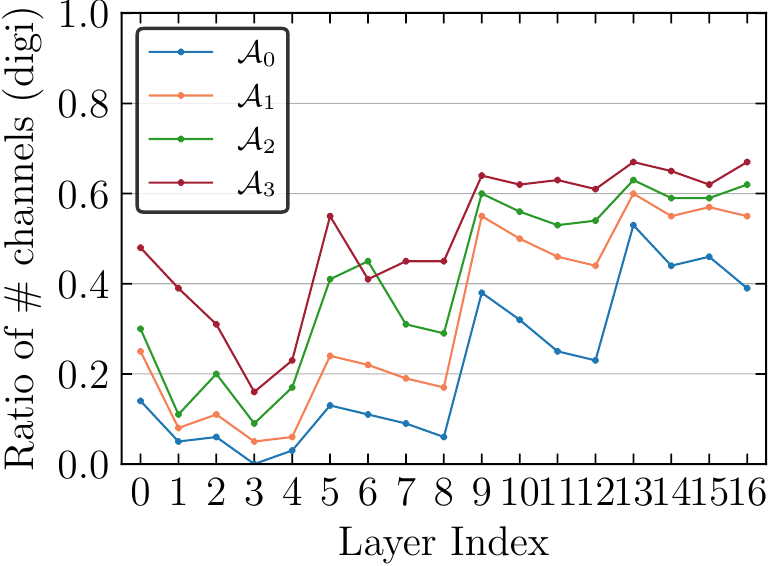}
    \label{fig:res18_adcR}
    }
    \vspace{-13pt}
    \caption{~\small
    The layer-wise ratio of channels assigned to the digital activation units~(\emph{digi}) on searched models.
    (a) VGG13 with MACAM-2 and ADC-2.
    (b) ResNet18 with MACAM-1 and ADC-1.
    }
    \vspace{-15pt}
\end{figure}

\noindent\textbf{Searched Layer-wise Assignment $\mathcal{A}$.}
We further validate the success of \name to learn a good assignment $\mathcal{A}$ for boosting expressiveness.
Fig.~\ref{fig:VGG13_adcR} and Fig.~\ref{fig:res18_adcR} visualize the ratio of feature map channels being assigned to digital activation path in each layer of VGG13 and ResNet18.
Under different activation energy cost constraints, \name learns a similar tendency to assign the workloads.
In former layers, the size of each channel of the feature map is larger, thus fewer channels are assigned to the electrical activation units to avoid violation of the energy constraint.
In contrast, more channels are assigned to the electrical activation units in the latter layers to boost the model accuracy.
This demonstrates our \name flow can automatically learn the optimized assignment with boosted expressiveness under energy constraints.

\begin{figure}
    \centering
    \vspace{-10pt}
    \subfloat[]{\includegraphics[width=0.20\textwidth]{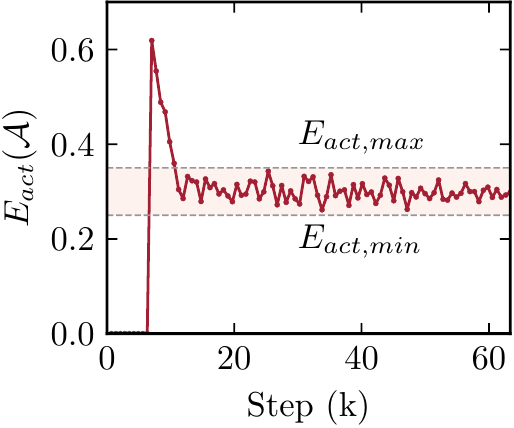}
    \label{fig:energy_curve}
    }
    \subfloat[]{\includegraphics[width=0.22\textwidth]{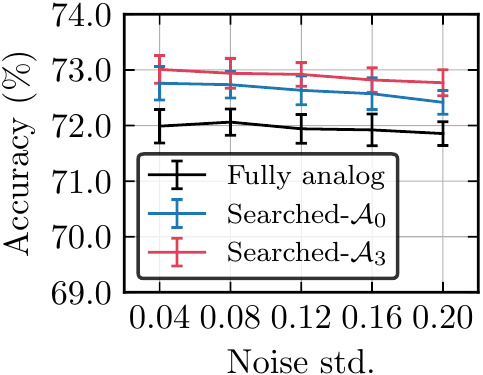}
    \label{fig:robustness}
    }
    \vspace{-13pt}
    \caption{~\small
    (a) Training curve of activation energy cost $E_{act}(\mathcal{A})$ of $\mathcal{A}_2$.
    (b) Robustness evaluation of Fully analog, $\mathcal{A}_{0}$, and $\mathcal{A}_{3}$· 
    Error bars show the $\pm 1\cdot \sigma$ variance over 20 runs.
    All models are VGG13 on CIFAR100 with MACAM-1 and ADC-1.
    }
    \vspace{-10pt}
\end{figure}

\noindent\textbf{Energy Penalty Curve}.~
In Fig.~\ref{fig:energy_curve}, the activation energy $E_{act}(\mathcal{A})$ is visualized along the \name training process.
$E_{act}(\mathcal{A})$ is well-bounded in the given constraint.
It continues the exploitation of the search space and converges with the temperate $\tau$ of the Gumbel Softmax approaching 0.

\noindent\textbf{Noise Robustness of Searched Models}.~
In Fig.~\ref{fig:robustness}, we evaluate the variation-robustness between searched models and model using fully analog activation units.
With the increasing variation on MACAM, our searched models show better noise
robustness since of the involvement of electrical activation units.

\vspace{-.05in}
\section{Conclusion}
\label{sec:Conclusion}
In this work, we propose a novel analog and \textit{mixed} activation system for energy-efficient neural network acceleration.
We first devise a \textit{fused} analog activation unit based on MACAM that is capable of achieving nonlinear ReLU-$\alpha$ and A/D conversion simultaneously, with superior energy efficiency to conventional digital activation implementation. 
We further integrate both the analog and digital activation dataflows to create a mixed activation system.
A \name training flow is developed to automatically learn how to assign activation workloads to the low-energy analog activation datapath and high-precision digital activation datapath, aiming at a balance of expressiveness and energy efficiency.
Our proposed methods are evaluated in a silicon photonic accelerator case study.
Compared to the standard activation implementation,
our mixed activation system with the searched assignment can achieve competitive accuracy with $>$60\% energy saving on the overall A/D conversion and activation energy cost.
Our MACAM-enabled analog and mixed activation system is viable to break through the curse of A/D conversion overhead in analog computing.

\section*{Acknowledgment}
\small  This work was supported in part by the Multidisciplinary University Research Initiative (MURI) program through the Air Force
Office of Scientific Research (AFOSR) contract No. FA 9550-17-1-
0071, and the Samsung GRO. 
The authors would like to thank Hao Chen, Mahshid Alamdar from The University of Texas at Austin and Xiyuan Tang from Peking University for helpful discussions.

\vspace{-.05in}
{
\footnotesize
\bibliographystyle{ACM-Reference-Format}
\bibliography{./ref/Top_sim,./ref/NN,./ref/NP,./ref/ALG, ./ref/HWA}
}

\end{document}